\documentstyle[11pt,psfig,paspconf,twoside]{article}
\input{gcbook.pgi}
\begin{document}

\title{Disks with Jet, ADAF, or EDAF for Sgr A$^{\star}$}

\author{A. C. Donea\altaffilmark{1}, H. Falcke\altaffilmark{2}
       and P.L. Biermann\altaffilmark{2}}
\affil{$^1$ Astronomical Institute of the Romanian Academy, Cutitul de Argint, 5, Bucharest, Romania}
\affil{$^2$ Max Planck Institute of Radioastronomy, Auf dem H\" ugel, 69, Bonn, Germany }

\begin{abstract}
We investigate various models of accretion disks for Sgr A*, one of
the most puzzling sources in the Galaxy. The generic image we have
taken into account consists of a black hole, an accretion disk, and a
jet. Various accretion models are able to explain the low NIR flux of
Sgr A*: a standard accretion disk with a jet, an ADAF, or an EDAF
(Ejection Dominated Accretion Flow) model. We find that all of these
models are conceptually similar. The accretion model which allows the
formation of the jet at the innermost edge of the disk requires a
sub-keplerian gas motion and a very large base of the jet. The large
base of the jet may be unrealistic for Sgr A*, since the jet model and
the observations suggest that the jet is collimated and anchored in
the disk in a very narrow region of the disk close to the black
hole. Alternatively, one can think of a jet plus wind model (EDAF), where
most of the energy goes out without being dissipated in the disk.
The model resembles the ADAF model at small radii. At large radii the energy is
ejected by a wind.

\end{abstract}


\keywords{accretion disk, jet, advection flow, black hole, EDAF,
ejection dominated accretion flow}


\section{Introduction}

As it is well known the Galactic Center harbors a black hole candidate
of $(2.5 \pm 0.4) \cdot 10^6 M_{\odot}$ mass (Eckart \& Genzel 1997).
The accretion rate inferred from Bondi-Hoyle accretion of stellar
winds should be small and should be of order $10^{-4}\dot {M_{\odot}}$
(Genzel et al. 1994, Melia 1992), but the exact number will depend
critically on the distribution of stellar wind sources in the Galactic
Center (Coker \& Melia 1999). What is peculiar about the central part
of our Galaxy is that, for a radiative efficiency of $10 \%$ of the
accretion flow, the accretion luminosity should be higher than
$10^{40}$ erg/s. This directly contradicts the bolometric luminosity
of $\sim10^{37}$ erg/s inferred from observations.  This means that a
standard accretion disk does not work.  To have such a low luminosity
the accretion disk in Sgr A$^\star$ has to be radiatively deficient.

There are several accretion models which in principle can produce such
 low luminosities: ADAF (advection dominated accretion flow) models
 (Narayan, Yi \& Mahadevan 1995, Narayan \& Yi 1994), disks driving
 jets from the innermost regions (Donea \& Biermann, 1996) or an EDAF
 (ejection dominated accretion flow) model (see below). All models
 provide a low luminosity which can explain the data of Sgr
 A$^{\star}$.

As it has been emphasized by Falcke \& Biermann (1995, 1999) a
symbiosis between the accretion disk, jet, and the central black hole
has become a widely accepted model for active galactic nuclei and
recently also for some of the microquasars.  The strong radio emission
in these systems comes from relativistic jets emitting synchrotron
radiation.  Based on the jet--disk symbiosis, Falcke (1999) has
explained the radio emission in Sgr A$^{\star}$ as being due to a
radio jet. The model fits well all current radio observations. In the
following we will make use of this model and the jet-disk symbiosis
idea to calculate a number of accretion disk models that would be 
consistent with the observed properties and upper limits of Sgr A*.

\section {Accretion Disk with Jet}

Donea \& Biermann (1996) have put forward an accretion disk model with a
jet starting at the inner regions of a disk. The jet extracts energy,
mass and angular momentum from small radii of the disk.  They have
written the conservation laws of mass and angular momentum taking into
account the extraction of mass, angular momentum and energy by the
jet (Newtonian treatment).  The new conservation mass law is:

\begin{equation}
\dot{M} = - 2 \pi R \Sigma u^{   R} + \dot{M}_{jet} 
\end{equation}
where $\dot{M}$ is the rate of mass accretion rate in the disk. The mass flow rate into the jet is $\dot M_{jet}$. The angular momentum conservation law  is:
 
\begin{equation}
2  \pi R \Sigma v_R(R)\   R^2 \Omega  = 2 \pi R \nu \Sigma R^2 \Omega^{'} + \dot {M}_{jet} v_{\phi}(R_{jet}) R_{jet} + C
\end{equation}
$\nu$ is the kinematic viscosity of the disk. $R_{jet}$ is the radius
where the base of the jet starts.  The constant $C$ is related to the
boundary conditions of the accretion flow (Frank et al. 1985).

We assume that the jet starts between $R_{\rm ms}$, the last
 marginally stable orbit in the absence of a jet and the radius
 $R_{\rm jet}$ with $R_{\rm jet}>R_{\rm ms}$. (Donea \& Biermann 1996).
 The presence of the jet will modify both the behaviour of the
 infalling matter across the radius $R_{\rm jet}$ and the structure of
 the relativistic disk. 

The local energy dissipation in the disk becomes:

\begin{equation}
D^*(R) = \frac{3 G M \dot M}{8 \pi R^3} \bigg[ (1-q_m) - (1-q_m) {\bigg( \frac{R_{jet}}{R} \bigg) }^{1/2}  \bigg]
\end{equation}
where 
\begin{equation} 
q_m = \frac{\dot{M}_{jet}}{\dot{M}},
\end{equation}
 As we mentioned before, the gravitational
 potential energy released between  $R_{\rm ms}$ and the outer
 radius of the jet, $R_{\rm jet}$ goes into the jet. Therefore, the
 total energy carried outwards by the jet  is strongly dependent on the
 mass and angular momentum of the black hole.  $ Q_{jet}$ is the total
 power of the jet -- including the rest mass energy of the expelled matter
 -- and is expressed as:

$$
Q_{jet} = L_{disk} - L_{disk}^{jet}
$$
$L_{disk}^{jet} $ is the total luminosity of a disk modified by the presence
of the jet and $L_{disk} $ is the total luminosity of the disk if
there are no physical conditions to drive the jet.

The power of the jet depends on the the way in which the jet gets
energy from the black hole/disk system.  Total power of the jet and
the radio activity of the AGN are probably directly connected to the
inner geometry of the system near the event horizon.  Adopting the
point of view that formation of jets is possible under all
circumstances (Falcke \& Biermann 1995), be it for an AGN or a stellar
size system, we investigate the effects of the presence of the jet on
the structure of the accretion disk, and implicitly on its emission
spectrum.

\begin{figure}[h]
\centerline{\psfig{figure=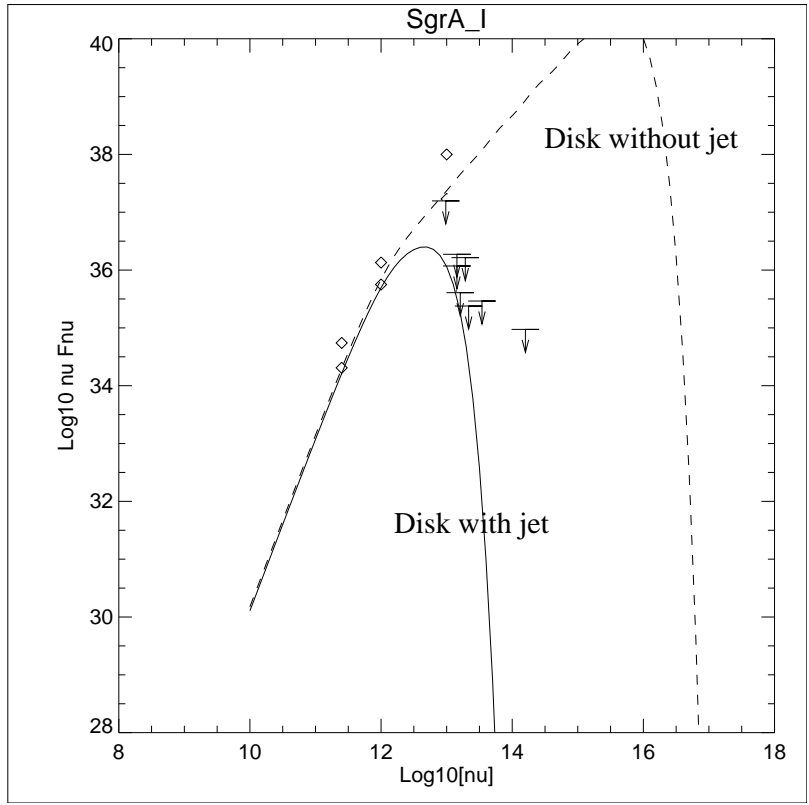,width=0.8\textwidth}}
\caption[]{
The emission from an accretion disk with a
jet starting at large radii in the disk.  $M = 2.5 \cdot 10^6
M_{\odot}$; Kerr black hole: $a_{\star}=0.9982$; $\dot M=2 \cdot
10^{-6} M_{\odot}/yr$; fitting the data below the NIR requires an
extremelysh  large base of the jet.} \label{fig-1} 
\end{figure}

 We applied the above disk--jet model to the case of Sgr A$^{\star}$.
 The spectrum from the disk is below the NIR flux if one considers the
 emission from a thin disk surrounding a Kerr black hole of $2.5 \pm
 0.4 \cdot 10^6 M_{\odot}$.  In Figure 1 we show a spectrum from a
 disk with jet (the dashed line) and without the jet (continuous
 line). As it is expected from the standard models (Shakura \& Sunyaev
 (1973), Novikov \& Thorne 1973) the standard model does not fit data
 at all.  If one takes into account the jet, then the model is
 completely different: the high energy part of the spectrum is cut off
 due to the extraction of energy by the jet from small radii.

   The parameters of the model (see Figure) are :\\

   -- low accretion rate: $\dot M=2 \cdot 10^{-6} M_{\odot}/yr$. \\

   -- black hole mass: $2.5 \cdot 10^6 M_{\odot}$.  \\

   -- Kerr black hole: $a_{\star}=0.9982$.     \\

 Interestingly, the NIR limits are now so stringent that fitting the
 data with such a model would require an extremely large base of the
 jet, from $ 1.23R_{\rm g}$ to $7000R_{\rm g}$.  The outer radius of
 the disk is $2\cdot10^5 R_{\rm g}$. The luminosity of the disk is
 $L_{\rm disk}=3.7 \cdot 10^{36}$ erg/s and $Q_{jet}$ takes out
 $99.5$\% of the total gravitational energy.

This is not a realistic situation for Sgr A$^{\star}$, since we know
that the radio emission is much more concentrated. Nevertheless, these
results point into two very interesting directions for modifying
accretion disk theory. For one, the disk model with large fraction of
energy extraction resembles an ADAF model (Narayan \& Yi, 1994) in
various points:

\begin{itemize}
\item[a)] the region between the radius of the last marginally stable
circular orbit $R_{ms}$ and the radius where we consider to be the
outer radius of the jet $R_{jet}$ no longer shows a standard disk-like
structure;

\item[b)] the flow becomes subkeplerian in the region where the jet
is formed;

\item[c)] at these radii there is no  radiative cooling any more of
the flow---the energy is advected in a jet instead of being radiated
away.
\end{itemize}

\section{The Ejection Dominated Accretion Flow}

A second way to think of our results, is that for such a large region
where energy extraction is required one can hardly speak of a 'jet'
any more. Rather, one could think of a continuous process, where a
magnetically driven wind across the accretion disk would extract a
large fraction of the dissipated energy. The jet would then only be
the most highly collimated and fastest feature in such a wind. 

Like in the ADAF model, the energy of the disk would no be radiated
away, but advected away. Unlike the ADAF model, however, the channel
through which the energy is advected would no longer be the disk, but
rather the wind. In light of these connections with the existing ADAF
theory we have labeled our approach EDAF---ejection dominated
accretion flow.

Indeed, the fact that accretion disks are able to expel a large
fraction by, most likely magnetically driven, jets opens up the
possibility that a related process could also drive large scale winds
from an accretion disk. Since it is often suggested that the viscosity
and dissipation in accretion disks is also a result of magnetic
fields, a deeper connection between viscosity, dissipation, and wind
ejection may exist. In cases where wind ejection dominates the energy
dissipation, a substantial part of the accretion energy could be
extracted by the wind from the accretion disk.  In this case the large
radius of $7000 R_{\rm g}$ we obtained earlier may be the radius where
a significant wind coexists with the disk (EDAF models). As pointed
out in Falcke (1999) application of the jet model indeed supports a
non-standard, radiatively deficient accretion disk in Sgr A$^\star$.
The EDAF model would be a situation where the wind goes off from the
accretion disk at every radius and where at each radius a fraction of
the available energy is put into a wind.

Falcke \& Melia (1997) solved the accretion disk equation for the case
of deposition of mass and angular momentum onto the disk by an
infalling wind. The reverse case, when the wind takes out mass and
angular momentum, would be an EDAF case. The equations remain the same, only
the sign of the mass loss/gain rate $\dot\Sigma_{\rm w}$ at each
radius changes.

We have used the code described in Falcke \& Melia (1997) to simulate
such a situation. The important parameter for these calculations is
the energy extraction efficiency: we assume that at each radius a
fraction $f_{\rm EDAF}$ of the locally dissipated energy is put into
kinetic and internal energy of the wind. The wind is assumed to leave
the disk with escape speed and with the specific angular momentum and
internal energy corresponding to the local conditions in the disk. The
local mass loss rate $\dot\Sigma_{\rm w}(r)$ is then completely
determined by $f_{\rm EDAF}$.

To facilitate the connection to the jet model we also assumed that the
wind extraction efficiency approaches unity at the inner $12 R_{\rm g}
$, i.e. all the energy is used to drive the jet modeled in Falcke
(1999). Because of the continuous mass loss throughout the disk, the
accretion rate at small radii and the available energy is
significantly reduced with respect to a standard disk model. Moreover,
the emitted radiation spectrum will also look markedly different
(i.e. flatter) from a standard $\alpha-$disk.

\begin{figure}[h]
\centerline{\psfig{figure=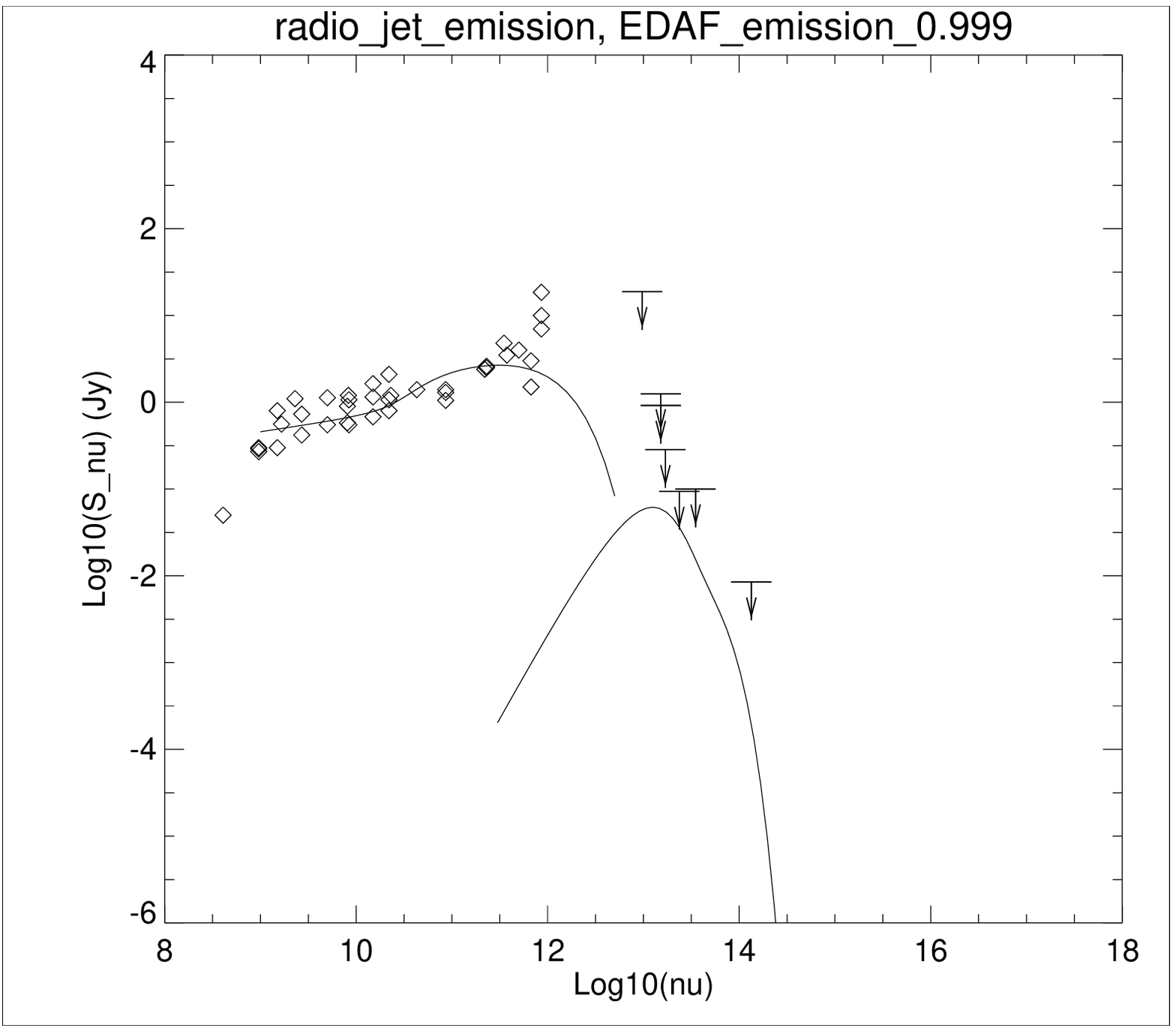,width=0.8\textwidth}}
\caption[]{ The radio emission and the spectrum from an
EDAF (Ejection Dominated Accretion Flow) model of Sgr A*.} \label{fig-2}
\end{figure}

The $F_{\nu}$ emission spectrum from an EDAF we have calculated for
the situation in Sgr A* is shown in Figure 2. The bump at low
frequencies is due to heating of the disk by the surrounding stars,
which we had to take into account as well. The spectrum lies below the
NIR limits. In Figure 2 the radio spectrum from the jet assumed to be
produced at the innermost radii by the remaining accretion flow is
also shown.  The great disadvantage of such a model is that $f_{\rm
EDAF}=99.995\%$ of the locally produced energy has to go into a wind
rather than into heat for an initial accretion rate of
$\dot{M}=10^{-6} M_{\odot}/$yr.  Even though we do not have any
arguments for how large the value of $f_{\rm EDAF}$ really has to be,
this seems uncomfortably high. On the other hand this high
'inefficiency factor' is very similar to those required for ADAF
models. In both cases one wonders whether such values are physically
meaningful. Perhaps various combinations of the models explored here
(e.g. ADAF+EDAF) could lead to a more convincing solution to the
inefficiency problem, or the mass accretion rate on Sgr A* is
significantly lower than the already low value used in our
calculations.

\section{Conclusions}

We conclude that there are in principle various ways to explain the
low NIR flux and that all the models (jet, EDAF, ADAF) are
conceptually very similar. However, each of these models seems to
require rather extreme parameters to explain the inefficiency of the
accretion flows.

The basic picture for the very central part of our Galaxy, explored
here, consists of an accretion disk, a black hole and a jet.  The
accretion disk may be an EDAF or a ``standard'' disk at large radii.
The angular momentum and part of the dissipated energy goes out with
the wind.  Perhaps at small radii, the disk becomes an ADAF with a
jet. The jet--disk model allows the formation of an jet at the
innermost radii and this would automatically produce a subkeplerian,
disipationless flow similar to an ADAF. In any case a large fraction
of the energy produced in the accretion disk in Sgr A* has to be
either advected into the black hole or ejected by a wind or jet.

\acknowledgments
A. C. Donea is grateful to Dr. Angela Cotera and Sera Markoff
for making her attendance to this conference possible. HF is supported
by DFG grant Fa 358/1-1\&2.

\end{document}